\documentclass[
 aip,
 amsmath,amssymb,
 reprint,
]{revtex4-1}

\usepackage{graphicx}
\usepackage{dcolumn}
\usepackage{bm}

\usepackage[utf8]{inputenc}
\usepackage[T1]{fontenc}
\usepackage{mathptmx}
\usepackage{etoolbox}

\usepackage{siunitx}
\usepackage{hyperref}
\usepackage{mhchem}

\makeatletter
\def\@email#1#2{%
 \endgroup
 \patchcmd{\titleblock@produce}
  {\frontmatter@RRAPformat}
  {\frontmatter@RRAPformat{\produce@RRAP{*#1\href{mailto:#2}{#2}}}\frontmatter@RRAPformat}
  {}{}
}%
\makeatother
\begin{document}

\preprint{AIP/123-QED}

\title[THz electrodynamics and superconducting energy scales of ZrN thin films]{THz electrodynamics and superconducting energy scales of ZrN thin films}

\author{Ozan Saritas}
\author{Frederik Bolle}
\author{Yayi Lin}
\author{Martin Dressel}
\affiliation{ 
{1.\  Physikalisches Institut}, Universit\"{a}t Stuttgart, 70569 Stuttgart, Germany
}

\author{Roman Potjan}
\author{Marcus Wislicenus}
\author{Andr\'{e} Reck}
\affiliation{%
Fraunhofer Institute for Photonic Microsystems (IPMS), Center Nanoelectronic Technologies (CNT), 01109 Dresden, Germany
}

\author{Marc Scheffler}
\email{marc.scheffler@pi1.uni-stuttgart.de}
\affiliation{ 
{1.\  Physikalisches Institut}, Universit\"{a}t Stuttgart, 70569 Stuttgart, Germany
}

\date{\today}

\begin{abstract}
The terahertz (THz) properties of ZrN thin films grown with CMOS-techniques on industry-standard \SI{300}{\milli\meter} silicon wafers are investigated in order to explore their superconducting behavior. The films have thicknesses ranging from \SIrange{18}{48}{\nano\meter}, and their critical temperatures $T_\textrm{c}$  are between 5 and 7.3~K. 
We probe the real and imaginary parts of the complex dynamical conductivity $\hat{\sigma}$ in the frequency range from \SIrange{100}{540}{\giga\hertz} (\SIrange{0.4}{2.2}{\milli\electronvolt}) and as a function of temperature. The experiments provide direct access to the low-energy electrodynamics and key materials parameters such as superconducting energy gap and superfluid density.
 Our findings indicate that ZrN is a weakly coupled BCS-type superconductor with a gap-to-$T_\textrm{c}$ ratio of $\approx 3.4$ in the thick film limit. For thinner films, this coupling ratio increases up to 4.0, departing from the BCS prediction. The results establish large-scale ZrN thin films as promising material for high-frequency superconducting applications.
\end{abstract}

\maketitle

The family of transition metal nitride superconductors continuously attracts research attention.\cite{Jena2019} Here one profits from different deposition techniques and from various tuning parameters that can be exploited during film growth, such as elemental composition, growth temperature, and film thickness. Nitride thin films can cover a large range of superconducting properties, from ultra-thin films\cite{Pracht2013} with reduced superconducting transition temperature $T_\mathrm{c}$, low superfluid density, and very high sheet kinetic inductance\cite{Bretz-Sullivan2022, TorrasColoma2024} to thicker films\cite{Horne2024} with $T_\mathrm{c}$ well beyond 10 K and high superfluid density. Other common advantages are high critical magnetic fields\cite{Baskaran2014} and robustness of thin films with respect to possible surface degradation\cite{Yadav2021} or multiple thermal cycling during cryogenic handling.\cite{Chen2021}

Research perspectives for superconducting nitride thin films thus are diverse: from a fundamental point of view, disordered nitrides are a major experimental testbed for physics near the disorder- or magnetic-field-driven superconductor-to-insulator transition.\cite{Breznay2017, Sacepe2020} 
They also enable superconducting nanowire single-photon detectors, which have reached commercial maturity and wide-spread use in quantum communication and optics.\cite{Koziy2021} 
Thicker films with high superfluid density are investigated for low-loss superconducting microwave circuitry and superconducting quantum bits.\cite{Barends2010, Kim2021} 

A rather recent direction of research concerning nitride superconductors is the preparation of thin films on larger wafers and with industry-standard (CMOS-compatible) fabrication techniques.\cite{Leon2021} 
One important step was the demonstration of HfN and ZrN thin films grown on \SI{300}{\milli\meter} silicon wafers.\cite{Potjan2023, Potjan2025} 
Along this line we study the THz properties of such ZrN thin films of different thickness, and thus we broaden the available information on superconducting ZrN thin films.\cite{Geerk1986, Tanabe1987, Cassinese2000, Zhu2007, May2024, Creswell2025, Kim2025} 
THz spectroscopy, giving direct access to key quantities of the superconducting state \cite{Dressel2008} such as the superconducting energy gap $\Delta$ and the superfluid density $n_\textrm{s}$,\cite{Dressel2008} has already been applied to other nitride superconductors.\cite{Gorshunov1993, Somal1996, Sindler2010, Beck2011, Henrich2012, Pracht2012, Hong2013, Pracht2013, Sherman2015, Simmendinger2016, Cheng2016} 
Our present experiments on the low-energy electrodynamics of ZrN thin films characterize their superconducting state and give an indication towards possible applications for high-frequency electronics or optics.

\begin{table*}
    \centering
        \caption{Characteristic material properties of superconducting ZrN samples: 
film thickness 
$d_\mathrm{TEM}$ observed in TEM,
 normal-state sheet resistance $R_\mathrm{s,DC}$,  normal-state low-frequency limit of THz conductivity $\sigma_\textrm{0,THz}$ at $T = 20\, \textrm{K}$, critical temperature $T_\mathrm{c}$ obtained from DC resistance measurements  ($T_\mathrm{c, DC}$) or from fits to THz data ($T_\mathrm{c, THz}$), energy gap $2\Delta_0 = 2\Delta(T=\SI{0}{\kelvin})$,  gap-to-$T_\mathrm{c}$ ratio $2\Delta_0/ (k_\mathrm{B} T_\mathrm{c,DC})$, superfluid density  $n_\mathrm{s,0} = n_\mathrm{s}(T = \SI{0}{\kelvin})$, London penetration depth $\lambda_0 = \lambda_\mathrm{L}(T= \SI{0}{\kelvin})$, sheet kinetic inductance $L^\square_\mathrm{kin,0} = L^\square_\mathrm{kin}(T= \SI{0}{\kelvin})$. The stated values of  $2\Delta_0$, $n_\mathrm{s,0}$,  $\lambda_0$, and $L_\mathrm{kin,0}$ are the zero-temperature extrapolations of the experimentally observed data.\\
}

    \begin{tabular}{c|c|c|c|c|c|c|c|c|c} \hline
         $d_\mathrm{TEM}$ (nm)& $R_\mathrm{s,DC}$ ($\Omega$/sq) & $\sigma_\textrm{0,THz}$ ($\Omega^{-1}\textrm{cm}^{-1}$)  & $T_\mathrm{c,DC}$ (K)&  $T_\mathrm{c,THz}$ (K)& $2\Delta_0$ $(\mathrm{meV})$&  $2\Delta_0/ (k_\mathrm{B} T_\mathrm{c,DC})$ & $n_\mathrm{s,0}$ ($10^{26}$ $  \mathrm{ m^{-3}}$) & $\lambda_\mathrm{0}$ (nm) & $L^\square_\mathrm{kin,0}$ (pH/sq)\\ \hline 
         17.6  &  64.47    &  9135  &   4.91    &  5.25     &  1.68   &    3.97  &   1.26  &  473  &  16.0  \\ \hline
         27.7  &  34.85    &  10076  &   6.35    &  6.30     &  1.96   &    3.58  &   1.65 &  414  &  7.76  \\ \hline
         36.4  &  24.02    &  10364  &   6.71    &  6.86     &  2.05   &    3.55  &   1.78  &  398  &  5.48  \\ \hline
         48.5  &  22.99    &  10783  &   7.29    &  7.37     &  2.15   &    3.42  &   1.98  &  378  &  3.70  \\ 
    \end{tabular}

    \label{tab:samples}
\end{table*}

\begin{figure}
    \centering
    \resizebox{1\linewidth}{!}{ 
    \begin{minipage}{0.499\textwidth}
        \centering
        \includegraphics[width=\textwidth]{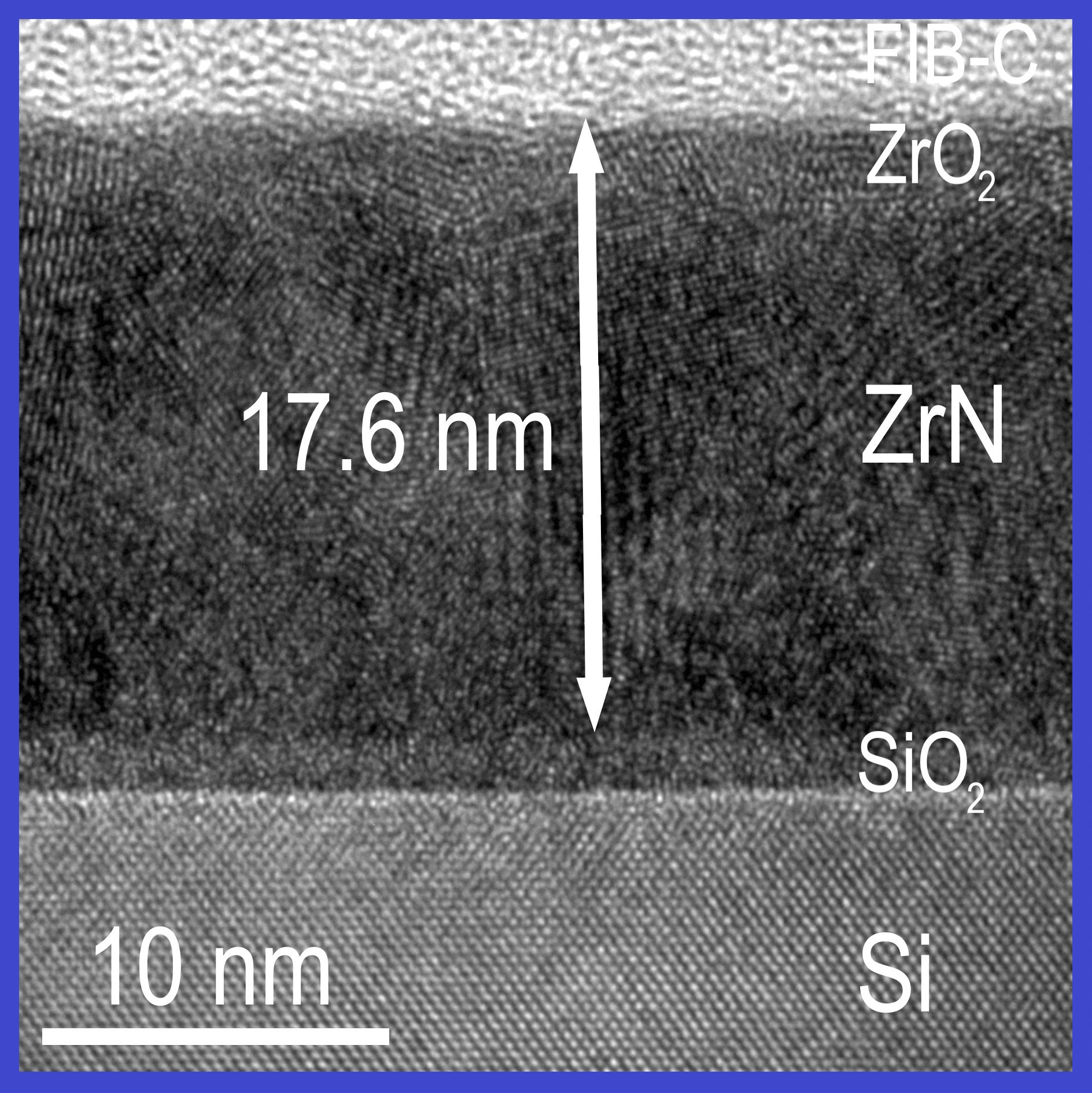} 
        \\
        \includegraphics[width=\textwidth]{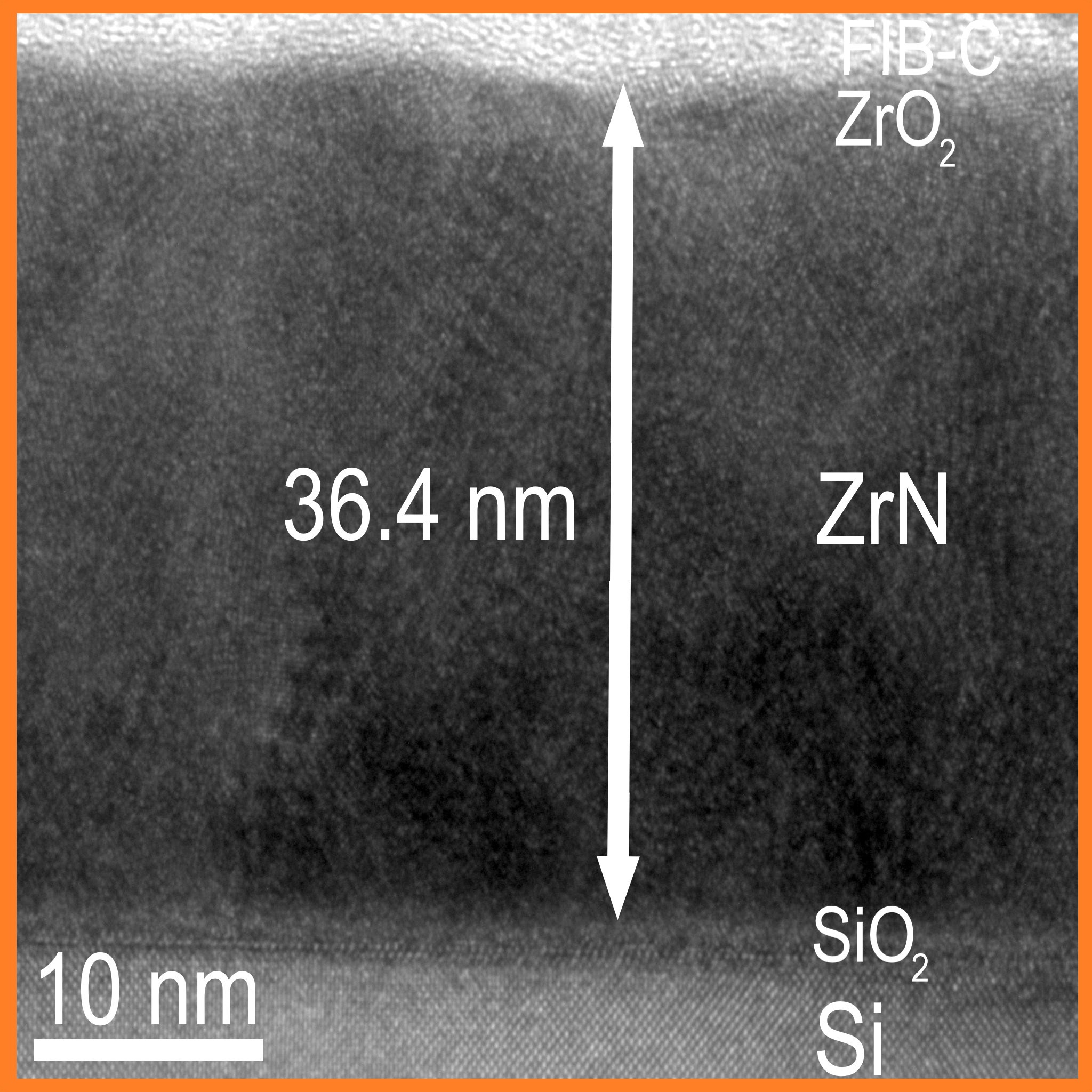} 
        \label{fig:firstcolumn}
    \end{minipage}
    \hfill
    \begin{minipage}{0.499\textwidth}
        \centering
        \includegraphics[width=\textwidth]{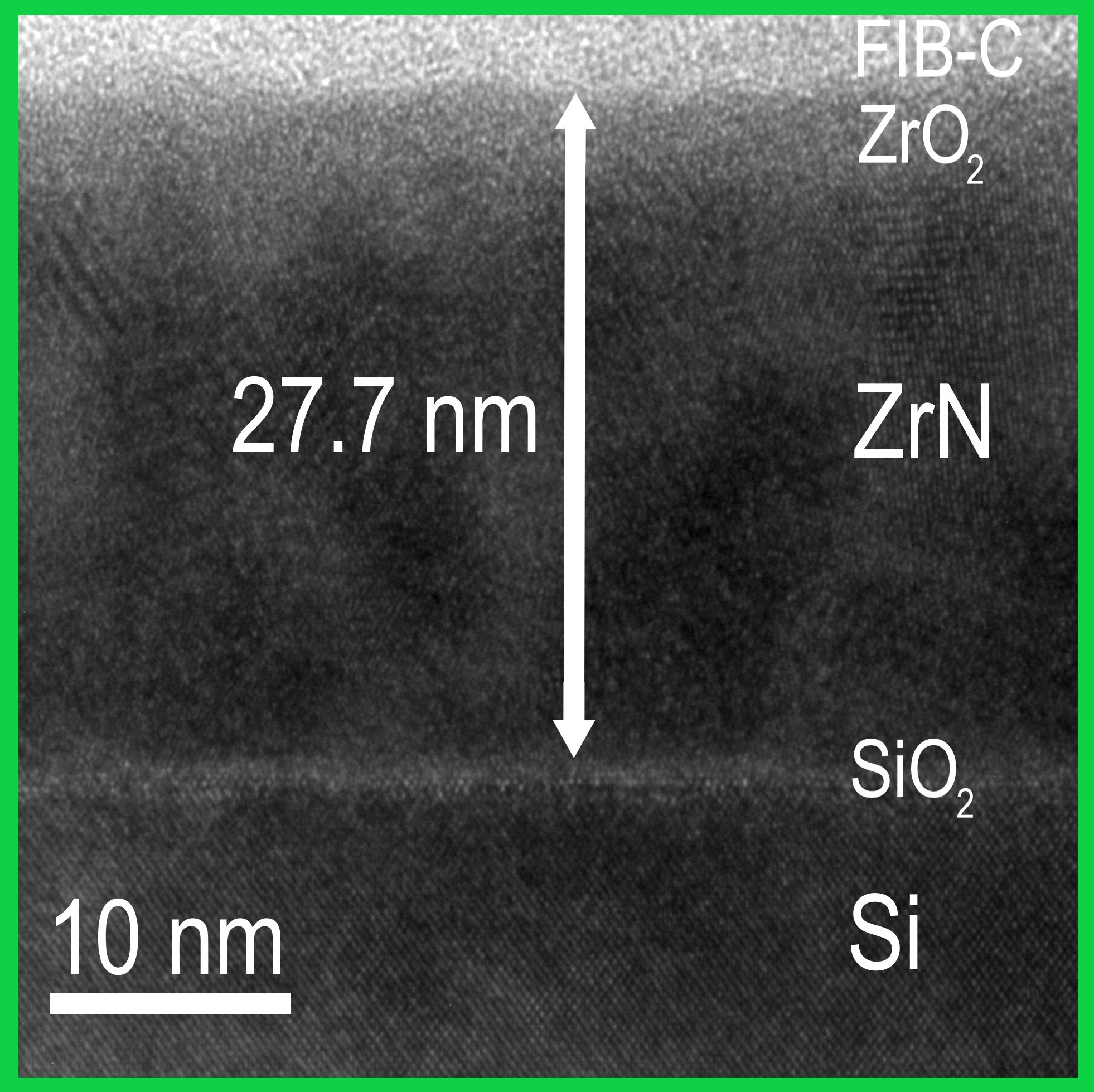}
        \\
        \includegraphics[width=\textwidth]{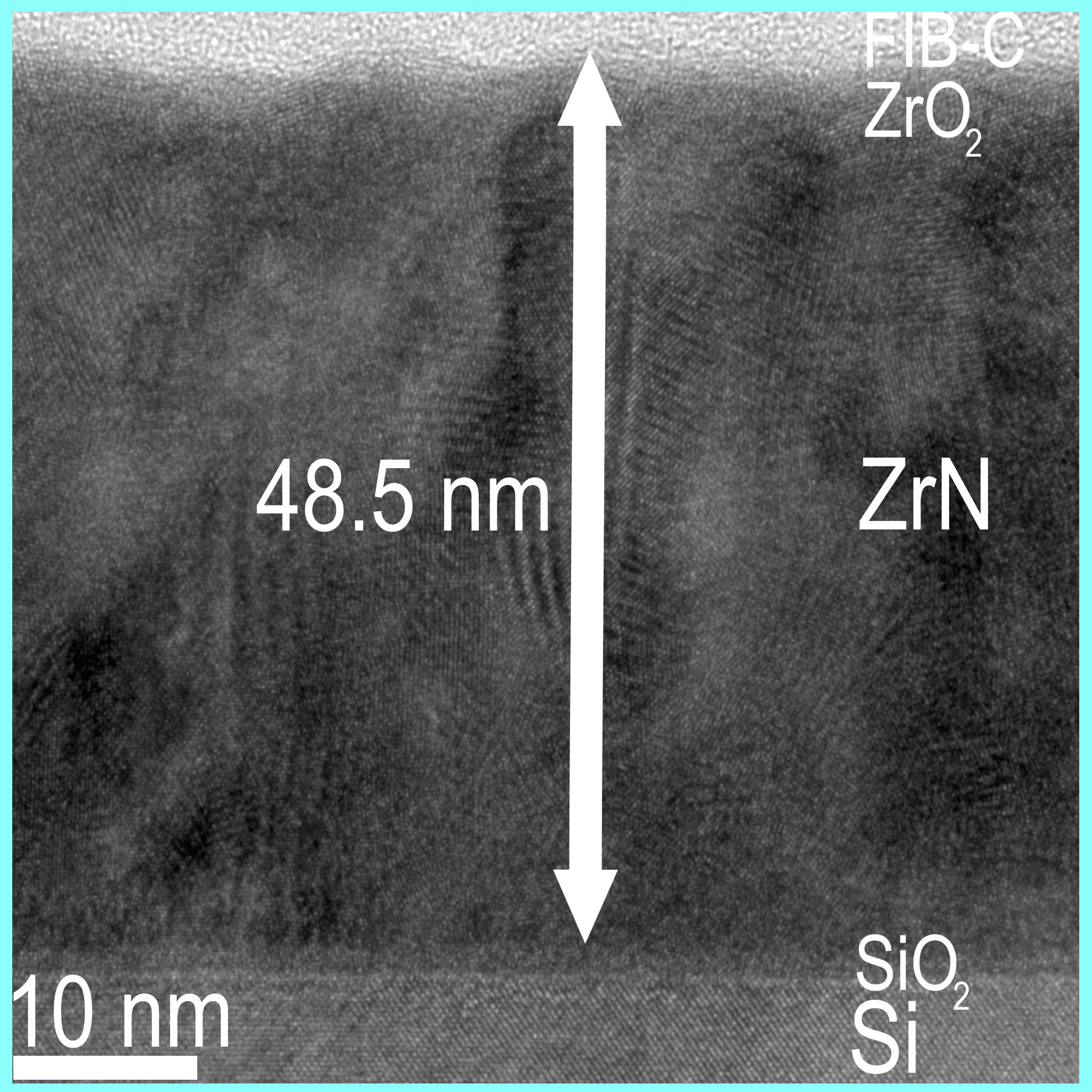} 
        \label{fig:secondcolumn}
    \end{minipage} }
    \caption{Bright field TEM pictures of the four samples with ZrN films with thicknesses of \SIrange{18}{48}{\nano\meter} on top of silicon substrates. Thin oxidation layers on the silicon wafer and on the ZrN films are also visible. }
    \label{fig:2x2grid}
    
\end{figure}

\begin{figure}
    \centering
    \includegraphics[width=1\linewidth]{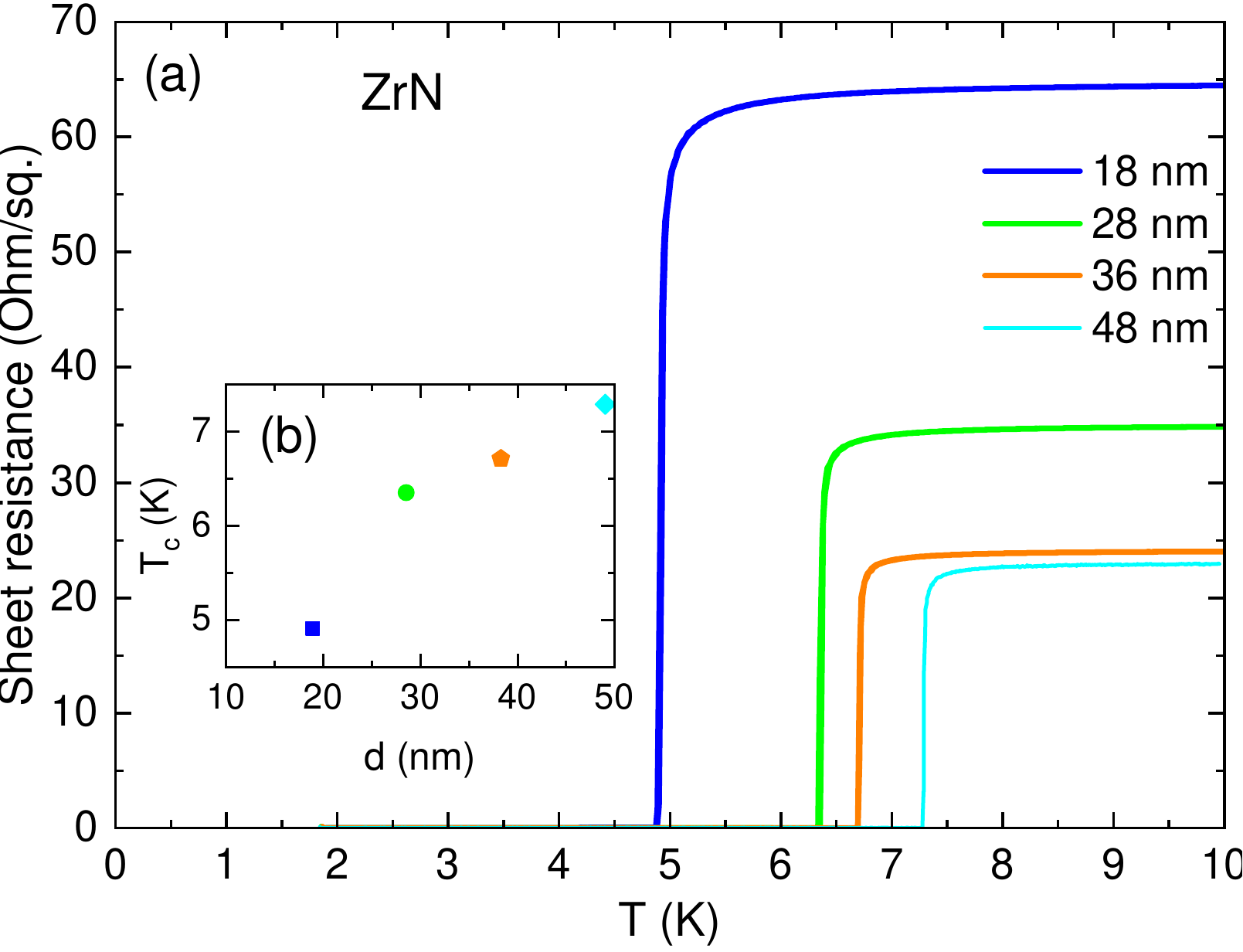}
    \caption{DC properties of ZrN thin films. (a) Sheet resistance as function of temperature for \SIrange{18}{48}{\nano\meter} thick ZrN samples, with sharp transitions to the superconducting state. (b) The inset displays the critical temperatures $T_\textrm{c}$ versus film thickness.}
    \label{fig:DC}
\end{figure}

The samples were fabricated on \SI{300}{\milli\meter} wafer-based process tools in an industry-standard, CMOS-compatible cleanroom.
The (100)-oriented, 300 mm high-resistivity ($\rho > 8000\ \Omega \text{cm}$) silicon wafers of thickness $775\,\mu\textrm{m}$ were cleaned with diluted hydrofluoric acid and degassed before deposition. 
ZrN thin films with a thickness of \SIlist[list-units=single,list-final-separator = {, and }]{18;28;36;48}{\nano\meter} 
were deposited by reactive magnetron sputtering in an Applied Materials Endura 2 cluster tool, equipped with a multicathode Clover PVD chamber.\cite{Potjan2023} 
Various properties of the four studied films are listed in \autoref{tab:samples}.

\autoref{fig:2x2grid} presents transmission electron microscope (TEM) bright-field images of the four ZrN samples. 
The ZrN films appear with a well-defined crystalline structure indicating a high quality and uniformity across the wafer.\cite{Potjan2023} 
The TEM pictures additionally reveal a thin oxidation layer (ZrO$_2$) on top of the metallization layer. Furthermore, the very thin native SiO$_2$ layer (about 1-2\,nm) can be recognized. Regarding the deposition roughness an increase can be observed with increasing layer thickness. Due to their dielectric nature and thickness of just a few nm, the thin oxide layers can be neglected for the analysis of the THz data presented below.

DC resistance data of the four samples are shown in \autoref{fig:DC}. The superconducting transition is clearly visible, with thinner samples exhibiting lower critical temperatures $T_\mathrm{c}$. The $T_\mathrm{c,DC}$ values as listed in \autoref{tab:samples} were determined by the 50\% criterion, i.e.\ $T_\mathrm{c,DC}$ is defined as the temperature $T$ where the DC resistance amounts to half the value of the normal-state resistance at temperature $T = \SI{10}{\kelvin}$.
The suppression of $T_\mathrm{c}$ with decreasing film thickness, explicitly shown in the inset of \autoref{fig:DC}, is frequently observed for nitride and other superconducting thin films.\cite{Semenov2009, Kamlapure2010, Ilin2012, Coumou2013, Ivry2014, Simmendinger2016, Haberkorn2018, Yadav2021, Bouteiller2025}

\begin{figure}
{\flushright
    \includegraphics[width=1\linewidth]{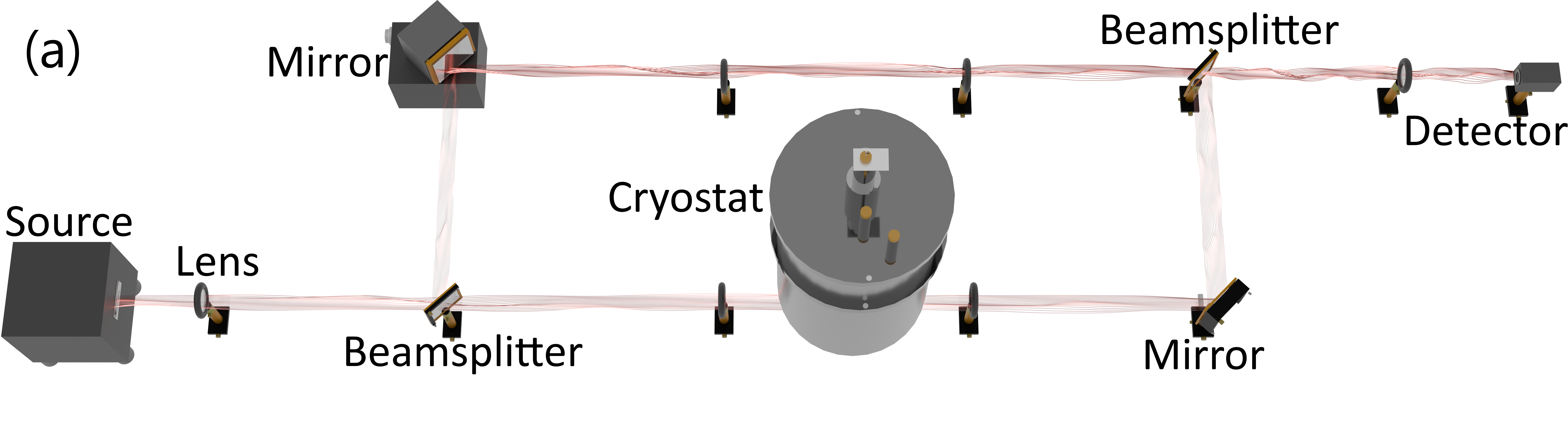}}
    \includegraphics[width=1\linewidth]{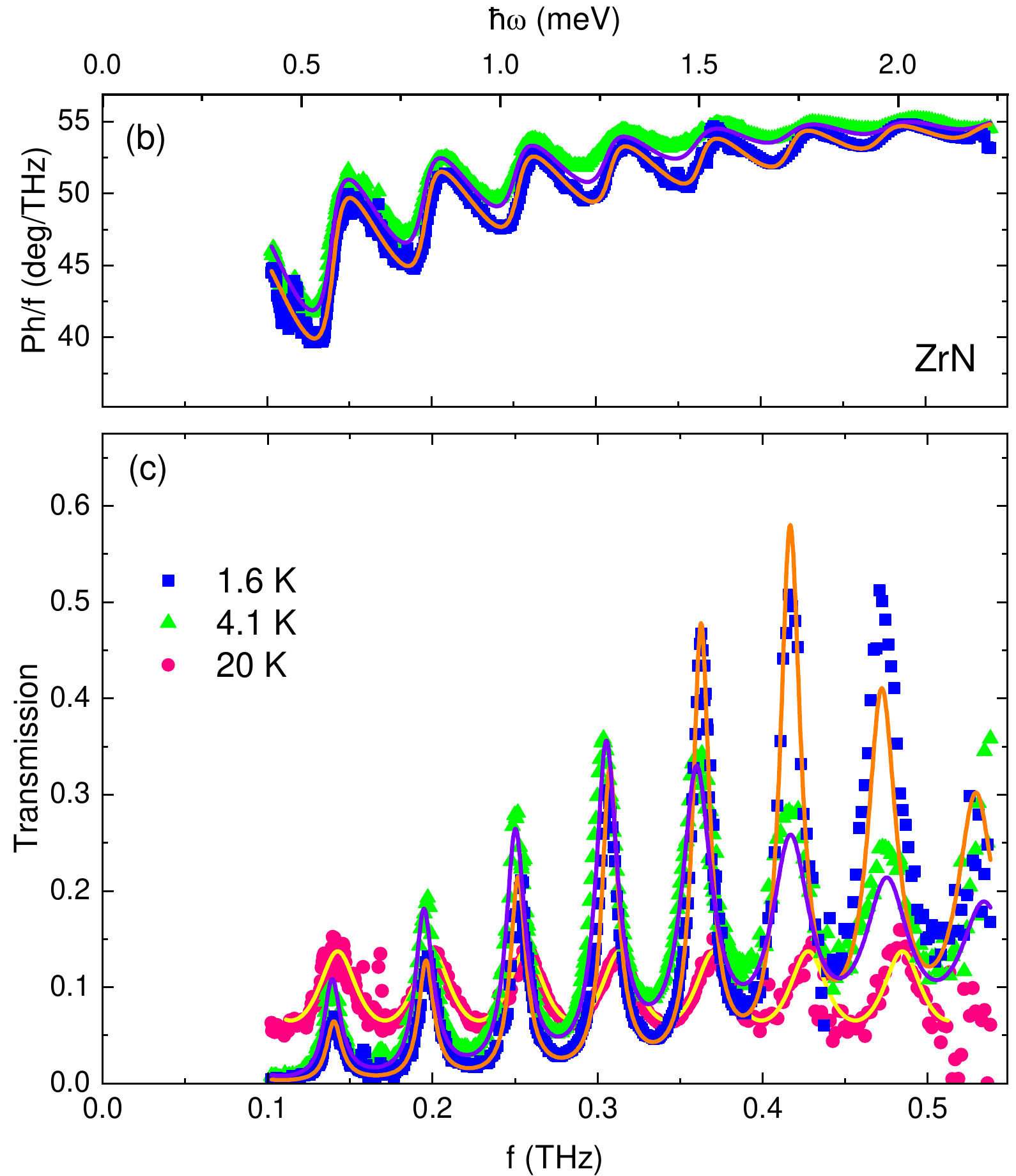}
    \caption{Setup and measurements of the \SI{18}{\nano\meter} ZrN sample for selected temperatures with associated fits, where $T_\mathrm{c} \approx \SI{5}{\kelvin}$. (a) THz frequency domain experimental setup. (b) Phase measurements and their associated phase fits. (c) Transmission measurements and their associated transmission fits, with additional data for a temperature above the superconducting transition temperature. }
    \label{fig:exp-setup}
\end{figure}

The main experimental data are acquired using a THz frequency-domain spectrometer in transmission geometry.\cite{Pracht2013} It employs backward wave oscillators (BWOs) as frequency-tunable source for monochromatic, coherent THz radiation, a Mach-Zehnder interferometer for phase-sensitive measurements, and a Golay cell for detection. The arrangement of these spectrometer components as well as additional elements like lenses and beamsplitters, is shown in \autoref{fig:exp-setup}(a).
The ZrN sample is cooled down using a home-built $^4$He-cryostat, which can reach temperatures down to \SI{1.5}{\kelvin} by pumping on the helium bath.\cite{Pracht2013}

For each measurement of the THz signal transmitted through a sample, a reference measurement of an empty aperture (i.e.\ sample removed from the THz beam path) is performed. Furthermore, the THz properties of a bare silicon wafer are also measured,\cite{vanExter1990, Hering2007, Pracht2012} which is needed for the quantitative analysis of the ZrN films on the silicon substrates.

The data analysis focuses on the complex transmission $\hat{t} = t\,  e^{(\mathrm{i}\phi)}$, where the absolute value $t$ governs the (power) transmittance as $|t|^2$, and $\phi$ is the phase shift introduced by the sample onto the transmitted THz wave.
To model these quantities, the transfer matrix method\cite{Dressel2002} is employed, where a two-layer system (vacuum, ZrN film, substrate, vacuum) is considered.
The resulting transmission coefficient can be represented as $t = t(\hat{\epsilon}_\mathrm{ZrN}, \hat{\epsilon}_\mathrm{sub}, d_\mathrm{ZrN}, d_\mathrm{sub})$. The thicknesses $d_\mathrm{ZrN} $ and $ d_\mathrm{sub}$ are already known, and the dielectric function $\hat{\epsilon}_\mathrm{sub}$ of the substrate is acquired with separate bare substrate measurements in the experiment. Therefore, the only unknown quantity is the dielectric function of the thin film, $\hat{\epsilon}_\mathrm{ZrN}$. 
The complex dielectric function $\hat{\epsilon}$ is directly connected to the complex optical conductivity $\hat{\sigma}$ via
\begin{equation}
    \hat{\epsilon}(\omega) = 1 + \mathrm{i} \frac{4 \pi \hat{\sigma}(\omega)}{\omega}
\end{equation}
resulting from the Maxwell equations.\cite{Dressel2002} Here $\omega = 2 \pi f$ is the angular frequency.
This expression connects the description of the complex THz transmission to the frequency-dependent complex optical conductivity $\hat{\sigma}(\omega)$, which is the response function that naturally describes the electrodynamical properties of mobile charge carriers of metals and superconductors.

As seen in the spectra shown in \autoref{fig:exp-setup}(b) and (c) we observe strong periodic oscillations in transmission and phase due to Fabry-P\'erot interference taking place inside of the silicon substrate, which acts as a dielectric slab with thickness comparable to the THz wavelength. 
In the normal state above $T_\textrm{c}$, the spacing of the oscillations  are mainly governed by the complex dielectric function $\hat{\epsilon}_\text{sub}$ and thickness $d_\text{sub}$ of the substrate layer while the ZrN thin films with its almost frequency-independent metallic conductivity determines the absolute transmission value at the Fabry-P\'erot maxima. Below $T_\textrm{c}$, the strongly frequency- and temperature-dependent transmission spectra are caused by the strong and characteristic changes in the optical conductivity $\hat{\sigma}_\text{ZrN}$ below $T_\textrm{c}$.

One approach to analyze the transmission data and to obtain absolute values for $\hat{\sigma}_\text{ZrN}$ is to fit the well separated, sharp Fabry-P\'erot peaks as distinct resonances, each with a characteristic frequency and width, and with $\hat{\sigma}=\sigma_1 + i\sigma$ as fit parameter for each resonance. Via this single-peak analysis \cite{Pracht2013, Dressel2008, Dressel2002} one obtains the frequency-dependent conductivity $\hat{\sigma}_\text{ZrN}$ from the THz spectra without explicit assumptions on the (super-)conducting properties of the ZrN thin films. 

Alternatively, we can assume that the underlying electrodynamics of the superconductor are described by conventional Bardeen-Cooper-Schrieffer (BCS) theory in the dirty limit. Such behavior is often found for nitride superconducting thin films,\cite{Pracht2013} and the results below confirm that such an assumption is justified also in the present case of ZrN.
In this approach, we utilize the Zimmermann equations,\cite{Zimmermann1991} which are an extension of the Mattis-Bardeen equations for superconductors of arbitrary purity, i.e.\ scattering processes are explicitly included concerning their influence on the superconducting electrodynamics below $T_\textrm{c}$. 
Then the optical conductivity depends on several parameters $\hat{\sigma}=\hat{\sigma}(\omega, T, \sigma_0,\Delta(T), T_\mathrm{c}, y)$, where $\sigma_0$ is the normal-state DC conductivity, $\Delta(T)$ is the temperature-dependent superconducting energy gap and $y= \hbar/(2 \tau \Delta)$ is a dimensionless "impurity parameter" characterizing the charge carrier scattering, with $\hbar$ the reduced Planck constant and $\tau$ the scattering time. 
Here $\omega$ and $T$ are parameters of the experiment, $T_\mathrm{c}$ and $\sigma_0$ can be acquired from the DC-measurements, and in our case $y$ can be safely set to $y=500$ far in the dirty-limit. 
Then only the temperature-dependent energy gap $2\Delta(T)$ remains unknown.
This allows for accurate fits of the frequency-dependent phase and transmission data, as shown for the examples in \autoref{fig:exp-setup}(b) and (c), where $2\Delta(T)$ as the only fit parameter is obtained for each temperature $T$. Here the fit includes the complete, broad spectral range of the measured data, and thus it is termed broadband analysis. 

In the following, we first discuss the THz conductivity $\hat{\sigma}(\omega, T)$ as key quantity to describe the THz electrodynamics of ZrN. From these data, we then obtain general parameters of the superconducting state, such as energy gap and superfluid density. Several of these quantities are listed in \autoref{tab:samples} for the four different ZrN samples.

\begin{figure}
    \centering
    \includegraphics[width=1\linewidth]{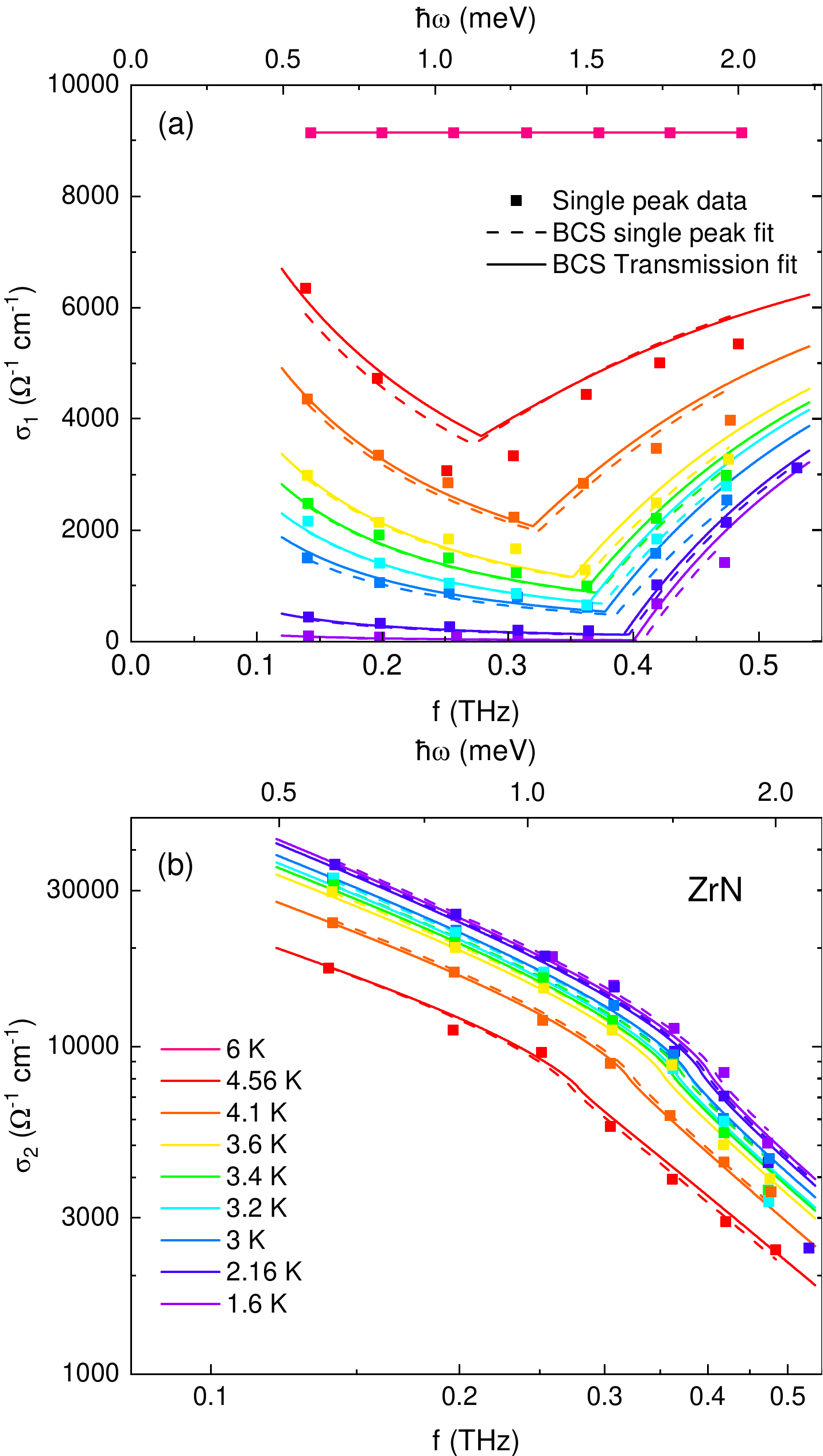}
    \caption{(a) Real part ${\sigma}_1(f)$ and (b) imaginary part ${\sigma}_2(f)$ of the frequency-dependent complex conductivity $\hat{\sigma}(f)$  of \SI{18}{\nano\meter} thick ZrN thin film, for different temperatures. Open symbols represent data obtained from single-peak analysis, which are then fitted, resulting in the dashed lines. The full lines are the conductivity spectra that correspond to fits to the measured transmission data, like those shown in  \autoref{fig:exp-setup}.
}
    \label{fig:opt-cond-comparison}
\end{figure}

Exemplary conductivity spectra for the 18~nm thick ZrN film are shown in \autoref{fig:opt-cond-comparison}. At $T = 6\,\textrm{K}$, in the metallic state, the real part $\sigma_1 \approx \sigma_0$ is basically frequency-independent (with values for $T= 20\, \textrm{K}$ for all samples listed in \autoref{tab:samples}), while the imaginary part $\sigma_2$ is neglible, as expected for a conventional metal in this frequency range.\cite{Dressel2006} Upon cooling below $T_\textrm{c}$, a minimum in $\sigma_1$ develops, which manifests the opening of the superconducting energy gap $2\Delta(T)$.
As expected, this minimum shifts to higher frequencies with decreasing temperatures.\cite{Pracht2013} Simultaneously, the inductive behavior of the superconductor develops, showing a clear $1/\omega$ dependence in $\sigma_2$ at frequencies below $2\Delta(T)$. Here $\sigma_2 \gg \sigma_1$, meaning that the lossless inductive response of the superfluid condensate dominates over the remaining lossy single-particle excitations, which are exponentially suppressed with decreasing temperature.

 \autoref{fig:opt-cond-comparison} shows two types of fits: dashed lines are the results of Zimmermann-type fits to the $\hat{\sigma}$ data points obtained via single-peak analysis. Full lines, on the other hand, are the conductivity spectra that correspond to broadband fits to the transmission spectra like those in \autoref{fig:exp-setup}. Clearly, these fits describe the experimental data well. This indicates that the superconducting electrodynamics of ZrN follows Mattis-Bardeen-type predictions, and thus that the superconducting state is properly modeled by BCS assumptions. 

\begin{figure}
    \centering
    \includegraphics[width=1\linewidth]{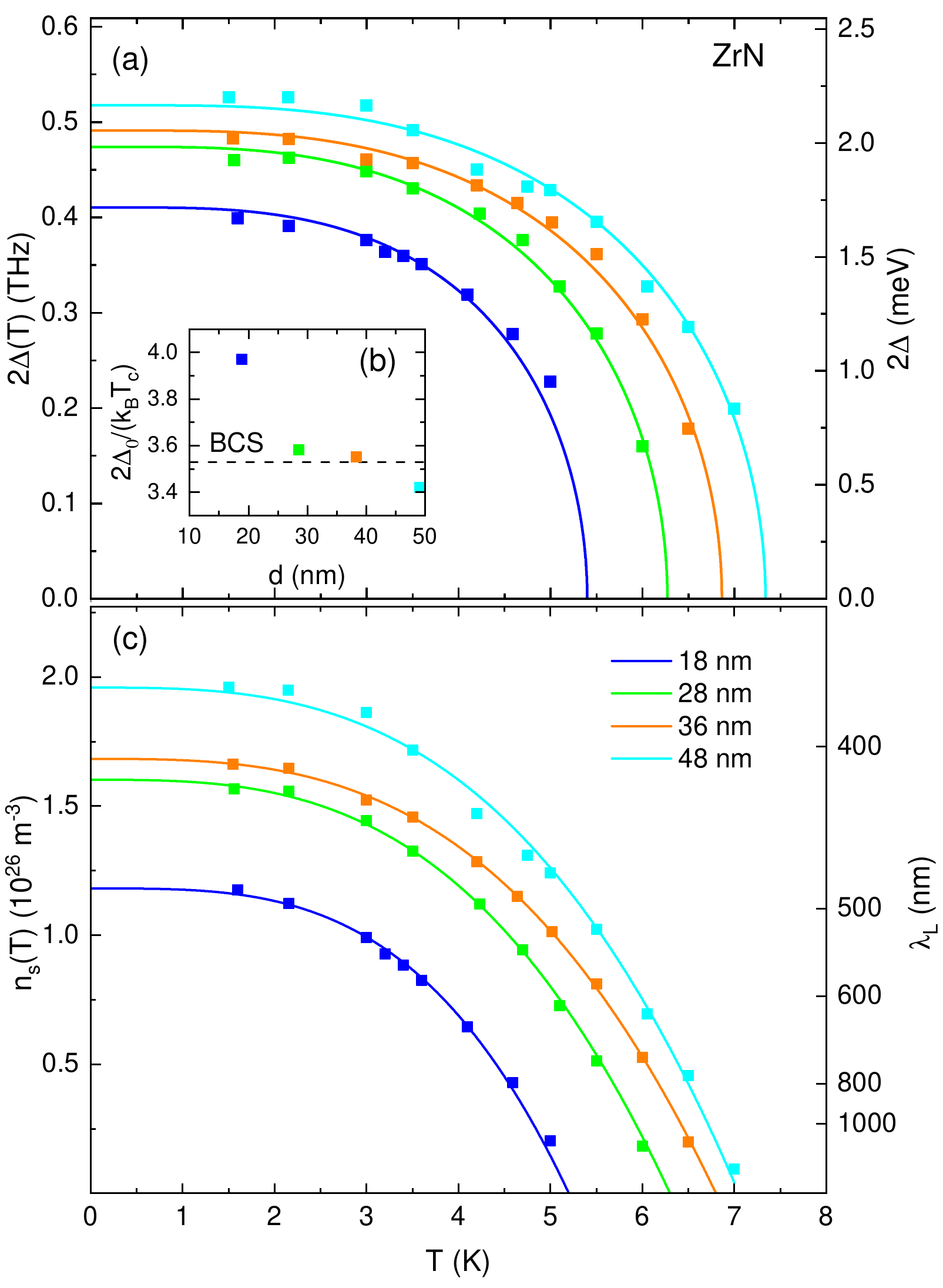}
    \caption{(a) Temperature-dependent superconducting energy gap $2\Delta(T)$ for ZrN films with thicknesses \SIrange{18}{48}{\nano\meter}. Fits follow  \autoref{eq:DeltaVsT}. Inset (b): Thickness-dependent gap ratio $2\Delta_0 / \mathrm{k_B} T_\mathrm{c}$, with the BCS prediction of 3.53 indicated as horizontal line. (c) Temperature-dependent superfluid density $n_\mathrm{s}(T)$ and London penetration depth $\lambda_\mathrm{L}(T)$ for the four ZrN samples of different thickness. Discrete $n_\mathrm{s}(T)$ points are acquired with direct broadband transmission fits. Lines are fits to a two-fluid-like model, \autoref{eq:two-fluid}. The right axis indicates the corresponding $\lambda_\mathrm{L}$. The zero-temperature limits of the fits are $\Delta_0$, $n_\textrm{s,0}$, and $\lambda_0$.}
    \label{fig:egap-superfl}
\end{figure}

The discrete points depicted in \autoref{fig:egap-superfl}(a) represent the energy gap values $2\Delta(T)$ obtained through direct broadband fitting of the transmission data. Comparing the four different samples, higher $T_\textrm{c}$ implies larger $2\Delta(T)$. Here the temperature dependence of the energy gap is is modeled using the formula
\begin{equation}\label{eq:DeltaVsT}
     2\Delta (T) =  2\Delta_0 \, \mathrm{tanh \Bigl ( \beta \sqrt{\frac{T_\mathrm{c,THz}}{T}-1}} \Bigr )
\end{equation}
to extract the energy gap $2 \Delta_0 = 2\Delta(T = \SI{0}{\kelvin})$ at zero temperature.\cite{Kajimura2012} 
The value of $\beta$ matches the BCS prediction, $\beta_\mathrm{BCS}= 1.74$, for all samples except the \SI{18}{\nano\meter} sample, for which $\beta= 2.0$.

Using the obtained fit parameters $\Delta_0$ and $T_\mathrm{c,DC}$ the gap ratio $2\Delta_0 / \mathrm{k_B} T_\mathrm{c,DC}$ can be calculated for each film thickness, as seen in the inset of \autoref{fig:egap-superfl}(a).
We observe a clear increase in the gap ratio from around 3.4 for the \SI{48}{\nano\meter} film up to 4.0 for the thinnest film studied in this work, indicating deviations from BCS theory, which predicts a value of 3.53 in the weak-coupling limit. Other works on disordered superconductors also report strongly disorder-dependent gap ratios such as in \ce{NbN}\cite{Chockalingam2009} and \ce{InO}\cite{Sacepe2011}. 
Such deviations naturally occur for bosonic-type superconductor-insulator transitions (SITs), where a pseudogap above $T_\textrm{c}$ emerges and an energy gap survives into the complete suppression of superconductivity beyond a critical value of disorder.\cite{Emery1995, Pracht2016}
It is difficult to draw the same conclusion for our ZrN data, as we did not investigate samples with even lower thickness and higher disorder, but the trend observed here could motivate further studies towards a possible SIT for strongly disordered ZrN. 

The superfluid density $n_\mathrm{s}(T)$ is directly connected to the imaginary part of the optical conductivity $\sigma_2$ through the Kramers-Kronig relations and the Ferrell-Glover-Tinkham sum rule \cite{Ferrell1958}
\begin{equation}
    n_\mathrm{s}(T) = \lim_{\omega \to 0} 4 \pi \omega \sigma_2(\omega, T).
\end{equation}
The superfluid density is calculated for each temperature point and is depicted in \autoref{fig:egap-superfl}(c). These data are then fitted using a modified two-fluid model
\begin{equation}\label{eq:two-fluid}
    n_\mathrm{s}(T) = n_\mathrm{s}(T = 0~K) \Bigl [   1- \Big(   \frac{T}{T_\mathrm{c}}  \Bigr)^c \Bigr],
\end{equation} to extract the temperature-dependent behavior.\cite{Poole2014}
Here the parameter $c$ in the canonical two-fluid-model is $c=4$. In our case the data can be fitted better with a value close to $c=3$, which aligns with prior findings, where other temperature dependences besides $T^4$ have been found.\cite{Hashimoto2013} 
The superfluid density allows for calculations of the London penetration depth\cite{Dressel2002}
\begin{equation}
    \lambda_\mathrm{L} =   \sqrt{\frac{m_e c^2}{4\pi n_\mathrm{s} e^2} },
\end{equation}
and the sheet kinetic inductance\cite{Meservey1969}
\begin{equation}
    L^\square_\mathrm{kin} = \frac{m_\mathrm{e}}{n_\mathrm{s}e^2} \frac{l}{A},
\end{equation}
where we assume $l/A = d_\mathrm{ZrN}$ for a square-shaped geometry.

The resulting zero-temperature limits of these quantities, $n_\textrm{s,0}$, $\lambda_0$, and $ L^\square_\mathrm{kin,0}$ are included in \autoref{tab:samples}.
The obtained values for $\lambda_0 \approx 400\, \textrm{nm}$ are in the range typical for thin films of conventional superconductors.
If one thinks towards the application of disordered superconductors in quantum circuitry, then the numbers for the sheet kinetic inductance, in the range of 10~pH/sq, are lower than values for low-superfluid-density superconductors like granular aluminum,\cite{Gruenhaupt2019, GlezerMoshe2020} but further reduction in film thickness, below the 18~nm studied here, could increase kinetic inductance substantially.

In conclusion, we have investigated the superconducting properties of four ZrN thin film samples of thicknesses \SIrange{18}{48}{\nano\meter} grown on \SI{300}{\milli\meter} silicon wafers.
THz spectroscopy provides access to key quantities such as the optical conductivity $\hat{\sigma}(\omega,T)$, the temperature-dependent superconducting energy gap $2\Delta(T)$, and the temperature-dependent superfluid density $n_\mathrm{s}(T)$. We find that ZrN mostly behaves as a conventional BCS-type superconductor and that it can be modeled accurately using Mattis-Bardeen equations. Superconductivity is suppressed as the film thickness is reduced, leading to a decrease in critical temperature, energy gap, and superfluid density. Furthermore, we report a significant increase in the coupling ratio for thinner films, which can be interpreted as the onset towards a disorder-induced superconductor-insulator transition. On the other hand, applications with thicker films, possibly enhanced $T_\textrm{c}$ and lower kinetic inductance for low-loss microwave circuitry also appear within reach.\cite{Cassinese2000}

The successful modeling of the high-frequency dynamical conductivity using the Mattis-Bardeen formalism allows accurate predictions regarding the quasiparticle losses and inductive response at GHz frequencies, which are important for designs of microwave devices and sensors.
Therefore, the present results demonstrate the potential of ZrN as a superconducting material that can be grown on industrial scales for various applications in quantum circuitry or cryogenic sensing.

This work was supported by the BMFTR within the project QSolid (FKZ: 13N16159) and by the Carl-Zeiss-Stiftung (QPhoton).
We thank Katrin Zimmermann for supporting the TEM sample preparation and analysis and Gabriele Untereiner for her support with sample preparation for THz measurements.

\section*{\label{sec:level1} References}
\nocite{*}
\bibliography{ZrN_Saritas_2025-09-20}

\end{document}